\documentclass[pre,aps,twocolumn,showpacs,floatfix,10pt,superscriptaddress]{revtex4-1}
\usepackage{graphicx}
\usepackage{color}
\usepackage{amsmath, amsthm, amssymb}
\usepackage{epsfig}
\newcommand{\be}{\begin{equation}}
\newcommand{\ee}{\end{equation}}	
\newcommand{\bea}{\begin{eqnarray}}
\newcommand{\eea}{\end{eqnarray}}

\begin{document}

\title{A class of exactly solved assisted hopping models of active-absorbing state transitions on a line}
\author{Rahul Dandekar}
\email{dandekar@theory.tifr.res.in}
\affiliation{Department of Theoretical Physics, Tata Institute of Fundamental Research, Homi Bhabha Road, Mumbai 400005, India}
\author{Deepak Dhar}
\email{ddhar@theory.tifr.res.in}
\affiliation{Department of Theoretical Physics, Tata Institute of Fundamental Research, Homi Bhabha Road, Mumbai 400005, India}

\date{\today}

\begin{abstract}
We construct a class of assisted hopping models in one dimension in which a  particle can move only if it does not lie in an otherwise empty interval of length greater than  $n+1$. We determine the exact steady state by a mapping to a gas of defects with only on-site interaction. We show that this system undergoes a phase transition as a function of the density $\rho$ of particles, from a low-density phase with all particles immobile for $\rho \le \rho_c = \frac{1}{n+1}$, to an active state for  $\rho > \rho_c$. The  mean fraction of movable particles in the active steady state varies as $(\rho - \rho_c)^{\beta}$,  for $\rho$ near  $\rho_c$.  We show that for the model with range $n$, the exponent $\beta =n$, and thus can be made arbitrarily large. 
\end{abstract}


\maketitle

{\bf Introduction} - There has recently been a lot of interest in  models showing an active-absorbing state transition \cite{grinstein,lubeckrev,hinrichsen}. The best studied of these  is the directed percolation (DP) class \cite{dp1,dp2}, and the related  pair-contact process with diffusion (PCPD) \cite{pcpd1,pcpd2,pcpd3} and parity-conserving directed percolation (PC) \cite{pc1,pc2}.  Models with conserved number of particles where a particle may hop only if there are sufficient number of other particles present within a given range, also show an active-absorbing phase transition, where the low-density phase is inactive. Examples of this class are the conserved lattice gas model and the conserved threshold transfer process\cite{rossi}, fixed energy sandpiles \cite{vespi} and the activated random walkers model \cite{arw}. By adding a slow drive and dissipation, these models can be  converted into models showing self-organized criticality \cite{brazilrev}.  It has recently been argued that on adding symmetry-breaking perturbations, these models would flow to DP-like critical behavior, but the evidence for this scenario is mainly numerical \cite{dharmohanty, mohantymanna}. It seems worthwhile to investigate  further models with conserved particle number showing active-absorbing phase transitions.\\

For critical phase transitions in non-equilibrium models, one can define critical exponents  in analogy with the equilibrium transitions. In particular, for the active-absorbing transitions, one defines the order parameter exponent $\beta$ by the condition that the mean activity in the active phase varies as $\epsilon^{\beta}$, where $\epsilon$ is a measure of distance from the critical point. For the DP, PC or PCPD classes, it is known from simulations that $\beta \leq 1$, in all dimensions where the transition exists. For the models with conserved particle number, the parameter to be tuned is the density, $\rho$. Models with $\beta>1$ are quite  untypical in  equilibrium or non-equilibrium  phase transitions. In fact, the only model showing an active-absorbing transition where $\beta$ is known to be greater than $1$ is the Model 3  defined by de Silva and de Oliveira in \cite{silvamj}. In this Letter we generalize this model and construct a class of models, with finite-range-interactions, for which $\beta$ can be determined exactly, and can be made arbitrarily large, taking all positive integral values.\\

Models of active-absorbing phase transition, with a conserved density, that have been solved exactly are restricted to  one dimension. The prototypical example of these models is the conserved lattice gas (CLG) model in one dimension, whose steady-state weights and critical properties have been exactly calculated (\cite{mario1d}\cite{mohantytr1}). There exist several variations on the basic CLG model (\cite{vanwijl}-\cite{jain}). All these models show a phase transition, as the density of particles is increased, from a phase with infinitely many absorbing states to a unique active steady-state.\\

One simplifying aspect of the models in this class is that, within the sector of steady-state configurations, the transition rules obey detailed balance such that all allowed configurations occur with the same probability. There is a mapping of the steady-state to the zero-range process \cite{mohantytr2},  and the calculation of correlation functions becomes very easy using this mapping. The models we study in this paper can be mapped to a gas of defects, but whose steady-state properties correspond to a zero-range process with a height constraint, and we use this to determine many steady-state properties of the system. \\

{\bf Definition of the models} - The models are defined on a line of L sites with N particles, with sites labelled $1,2, \ldots L$. At most one particle can occupy one site.   A configuration is denoted by a binary string $0^{m_1}$1$^{n_1}$0$^{m_2}$..., where 1 denotes a particle and 0 denotes an empty site. We call a connected cluster of empty sites as a 0-cluster (of the given length) and a connected cluster of occupied sites as a 1-cluster.   Particle can hop by one or more steps across empty regions, and cannot cross each other.  The model with range n is defined by the following two types of hopping processes:\\

a) The first  process is a particle hop of length at most n, activated by a nearest neighbour occupied site. A hop of length $i$ from a 1-cluster of length $>1$ into a 0-cluster of length $k$ takes place with rate $\Gamma_1(i,k)$,  to the right  or towards the left. This is represented by the processes  for all $i  \leq k$, and $m \geq 1$:

\bea
1^{m+1}0^k &\overset{\Gamma_1(i,k)}{\rightarrow}& 1^m0^i10^{k-i},  \\
0^k1^{m+1} &\overset{\Gamma_1(i,k)}{\rightarrow}& 0^{k-i}10^i1^m,  
\eea
For $i < k$, this leads to break-up of $0$-cluster length $k$ into clusters of length $i$ and $k-i$. 
We assume that for all  $k>n$, there is at least one $i<k$ for which $\Gamma_1(i,k)$  is non-zero. Thus all $0$-clusters of length $>n$ are susceptible to break-up into smaller clusters. \\

b) The second  process is the coalescence of 0-clusters separated by a single particle, where the resulting cluster has total length $ \leq n$. The single particle between the two 0-clusters  of length $i$ and $i'$ is transferred either to the nearest $1$ on the left with rate  $\Gamma_2(i, i+i')$ or to the right with rate $\Gamma_2(i',i+ i')$. This leads to joining of the 0-clusters into a single cluster of length $i+i'$, and  is represented by the processes : For  all $ m, m' \geq 1$, and  all $i, i'$, with $i + i' \leq n$

\bea
1^{m}0^{i}10^{i'}1^{m'} &\overset{\Gamma_2(i,i+i')}{\rightarrow}& 1^{m+1}0^{i+ i'}1^{m'},  \\
1^{m}0^{i}10^{i'}1^{m'} &\overset{\Gamma_2(i',i+i')}{\rightarrow}& 1^{m}0^{i + i'}1^{m'+1}, 
\eea

All hopping rates $\Gamma_{2}(i,k)$  are  non-zero for all $ i \leq k \leq n$, and $\Gamma_2(i,k) =0$ if $k> n$. \\

Finally, we state the conditions for detailed balance to hold in the steady-state sector. For any hopping process that does not lead to a breakup of $0$-cluster of length $> n$, the reverse hop occurs with equal rate.  Thus, we  assume for all $ i \leq k \leq n$

\bea
\Gamma_{1}(i,k) = \Gamma_{2}(i, k) {\rm ~for ~all~} 1 \leq i \leq k \leq n.
\eea

{\bf Characterizing the steady state} - For the model with range n, the break-up of 0-clusters of length $>n$ due to processes $\Gamma_1$  is irreversible, but the break-up of 0-clusters of length $\le n$ is reversible due to the processes $\Gamma_2$. Thus the number of clusters of length $> n$ can only decrease with time.  Configurations with 0-clusters of length $>n$ are thus either transient or absorbing, and cannot be part of an active steady-state. As a result, for $\rho \le 1/(n+1)$, the system always ends up in an absorbing state after passing through a series of transient configurations. The number of absorbing configurations increases exponentially with the size of the system. For $\rho > 1/(n+1)$, eventually, all 0-clusters that remain are of length $ \leq n$. These keep on breaking up, and reforming, as the particles hop, but always remain of length $\leq n$. Thus the critical density for the model with range n is $\rho_c = 1/(n+1)$. \\

For models with range n$>2$, there also exist absorbing states for densities above $\rho_c$. For example, consider the configuration $10^{r_1}10^{r_2}10^{r_3}1\ldots$, where $r_i$  are any integers with $r_i + r_{i+1} \geq n+1$, for all $i$. Here,  clearly, all $1$'s are immobile, and the maximum possible density possible in such a configuration is  $\rho_{abs}^{max} = 2/(n+3)$. \\

Now consider a configuration with density $\rho$, with $\rho_c \le \rho \le \rho_{abs}^{max}$, which has at least one mobile particle. As the particle moves, it may give rise to a configuration in which all particles are immobile. In such a case the previous configuration would have a larger $0$-cluster of length $ > n$. Such a configuration would be a transient configuration. The other case is when there are no $0$-clusters of length $>n$. In such a configuration, for all allowed hoppings, the reverse move is also allowed. Also, from any such active configuration $C$ we can go to a standard configuration $C_0$, in which all the number of $0$-clusters is the least possible (i.e. $\lceil (L-N)/n \rceil$), and they occur as far to the right as possible. Then, since  in this subset of configurations all moves are reversible, we can go from any configuration to any other, the active steady state is unique. Also, using the fact that detailed balance condition holds for reversible moves, all recurrent configurations  in the steady state occur  with equal weight. \\

The case $n=1$ is the CLG model. Here the only allowed  transitions are $110 \rightarrow 101$ and $011 \rightarrow 101$, both occuring with equal rate. Here $\rho_c = \frac{1}{2}$, and there are no absorbing states above $\rho_c$. The steady-state consists of clusters of 1s separated by single 0s (e.g.  $0110111110110...$). The case $n= 2$ has been studied as Model 3 in \cite{silvamj}. In this case $\rho_c = \frac{1}{3}$, but there exist absorbing states even above $\rho_c$, for eg. the absorbing state with the maximum density 1001010010... for which $\rho = 2/5$.\\

In the steady-state, all configurations with $N$ particles  on $L$ sites with no 0-clusters of length $>n$, and having at least one active site,  are present with equal probability. Let $C(N,L)$ is the number of such configurations with N particles on an open line of $L$ sites on a line, that start with a 1. We define

\bea
C_L(x) &=& \displaystyle\sum_{N=1}^{L} C(N,L) x^N  \label{eq:genfunc} \mbox{\quad\quad and}\\
C(x,y) &=& \sum_{L=1}^{\infty} C_L(x) y^L
\eea

If we include in this sum configurations without any active particle, It is easy to see that all such configurations are generated by the strings 1, 10, 100, ..., 10$^n$. The error incurred due to this over-counting decreases exponentially with $L$. Then,

\be
C(x,y) \approx \left(1-xy\frac{1-y^{n+1}}{1-y}\right)^{-1} - 1 \label{eq:genf}
\ee

For fixed finite $x$, $C_L(x) \sim \Lambda(x)^L$, where $\Lambda(x) $ is the largest solution of

\bea
\Lambda^{n+1} = x \frac{1-\Lambda^{n+1}}{1-\Lambda}\label{eq:Lambda}
\eea

For $x$ tending to $0$, this equation may be solved to give
\begin{equation}
\Lambda_+(x) = x^{1/(n+1)}\left(1+\frac{1}{1+n}x^{1/(n+1)} + \cdots\right)
\label{eq:Lambda}
\end{equation}

Using  $\rho = x\frac{d}{dx}\log{\Lambda(x)}$, we get 
\bea
\frac{1}{\rho} &=& (n+1) - \frac{\Lambda}{1-\Lambda} + \frac{(n+1)\Lambda^{n+1}}{1-\Lambda^{n+1}}
\label{eq:rho}
\eea
this gives
\bea
\rho  &=& \frac{1}{n+1} \left(1+\frac{1}{1+n}x^{1/(n+1)} + \cdots \right)
\eea

We can also determine the second largest root of Eq.(\ref{eq:Lambda}). this gives

\begin{equation}
\Lambda_-(x) = \omega x^{1/(n+1)}\left(1+\omega \frac{1}{1+n}x^{1/(n+1)} + \cdots\right)
\end{equation}

where $\omega= \exp[2 \pi i/(n+1)]$. The particle-particle correlation length is given by

\bea
\nonumber
\xi^{-1}(x) &=& \log{\left\lvert\frac{\Lambda_+(x)}{\Lambda_-(x)}\right\rvert} \\
 &\sim& x^{1/(n+1)} \label{eq:xil}
\eea

Denoting $\rho-\rho_c$ by $\epsilon$, we see that $\epsilon \sim x^{1/(n+1)} \sim 1/\xi$. We define $\nu$ by the equation $\xi \sim \epsilon^{-\nu}$. This gives us, for all $n$, $\nu =1$.\\

The reason for the appearance of $x^{1/(n+1)}$ in the above equations is not immediately obvious. In the following, we present a mapping to a gas of  defects and anti-defects, which makes it so.\\

{\bf Mapping to a Gas of Defects} - Consider the model with range n. Here the asymptotic state at the  critical  density is the inactive periodic configuration $10^n10^n10^n\ldots$( or its translates). We take this to be the reference configuration. A general configuration can be represented as $10^{s_1}10^{s_2}10^{s_3} \ldots$ where $s_i$'s give the lengths of the 0-clusters. We write $r_i = n-s_i$. A $0$-cluster of length $n -r_i$ will be said to have $r_i$ defects. Here $r_i$ can be negative, in which case, the cluster is said to have $s_i - n$ anti-defects. A defect is denoted by $x$, and an anti-defect by $\bar{x}$.  Then a general configuration with $N$ occupied sites can be represented by a gas of defects and antidefects on $N$ sites with $r_i$ defects at site $i$. We will denote a configuration, say with $( r_1, r_2,r_3, \ldots) =( 2,1,0,-1,3, \ldots)$ as $1xx1x11\bar{x}1xxx\ldots$, where the 1's are now considered as sites and the $x$'s are particles on the 1 to the left of them.\\

The transition rules (1)-(4) translate into the rule that two neighbouring sites can exchange defects at finite (given) rates if they have a total of $>$n defects. Thus, under time evolution, $\{r_i\}$ evolve by a groups of defects together jumping to a neighboring site to the left or right. However, the $r_i$'s are always constrained to be $\le$ n. Antidefects do not move. If a defect and anti-defect meet, they annihilate each other. For densities $\rho <\rho_c$, the final state has no mobile defects. For $\rho > \rho_c$, in any recurrent configuration with non-zero activity, there are no antidefects left.\\

For a given value of $N$ and $L$, all configurations in the steady-state are equally likely. The static properties in the {\em active} steady state can thus be calculated by considering number of the configurations $C_d(N_d,N)$ of a gas of  nearly  free defects, and no anti-defects. The only constraint is  that no site can have more than $n$ defects. For low densities, the gas of defects is nearly an ideal gas. Adding one particle is equivalent to adding $(n+1)$ defects. These defects unbind, and at low densities, form a nearly ideal gas. The fugacity associated with one defect is $x^{1/(n+1)}$, and forms a natural variable to describe properties of this gas.\\

 Since all configurations with a fixed number $N_d = (n+1)N-L$ defects on $N$ sites are equally likely, we can define the generating function $C_d(w,z) = \sum_{N_d=0,N=1} C_d(N_d,N) w^{N_d} z^N$. It can be verified that $C_d(w,z) = C(x=zw^n, y=1/w)$. \\

In terms of defects, the activity is the probability that two neighbouring sites have a total of $\geq n$, but strictly less than $2n$ defects. (If two neighbouring sites have $2n$ defects then the original configuration is 111, and hence the middle particle cannot move.) This probability can be calculated to give (using the fact that in the thermodynamic limit $w = 1/y = \Lambda_+$)

\be
\rho_a = \rho \Lambda^n \frac{(n+1) - (n+2)\Lambda - \Lambda^n + 2\Lambda^{n+1}}{(1 - \Lambda^{n+1})^2}
\ee

Near the transition, since the gas of defects is nearly ideal, the probability of two neighbouring sites having a total of $\geq n$ defects varies as $n$-th power of density of defects, hence as $\epsilon^n$. As the exponent $\beta$ is defined as mean activity varies as $(\rho -\rho_c)^{\beta}$, this immediately gives us that $\beta =n$.\\

If we study the decay of activity with time, when the initial state has exactly critical density, then this process can be expressed as the annihilation dynamics of defects and antidefects, where the antidefects do not move, but are eventually annihilated by invading defects. Thus  the activity at the critical density would decay as $t^{-\frac{1}{2z}}$ where $z$ is the dynamical exponent in the steady-state. This has been observed in the CLG (which is the model with range 1) \cite{leelee1} and has also been explained in terms of a mapping to a two-species annihilation process \cite{jain}.\\

{\bf Summary and Concluding Remarks} - In summary, we have studied a class of non-equilibrium models  of assisted hoppings in one dimension, characterized by a range parameter $n$, for all positive integer values of $n$. These are a generalization of the $n=2$ model studied by De Silva and De Oliveira \cite{silvamj} to general $n$.   For $n >1$, they have  absorbing states even for densities above the critical density. We determined the steady-state measure exactly and showed that the critical exponents have the values $\beta =n $ and $\nu =1$ for these models.\\

It is straight-forward to extend our discussion to models with asymmetrical hoppings, where particles can  hop  only to the right (say), on  a ring. For concreteness we set all non-zero rates to 1. Then, one can check that in the active state, all recurrent configurations occur with equal probability. There is a net mean current in the active steady state, which is proportional to mean activity, and  also varies as $|\rho - \rho_c|^{\beta}$, with $\beta =n$. \\

At the critical density, one expects a power-law decay of the activity with time, but this exponent is known to depend on the initial state \cite{mohantymanna}. In these models, one can characterize the structure of transient and steady state configurations in some detail. Adding a very small dissipation and drive, the self-organized version of these models can be studied. We plan to discuss these in a future publication.

\begin{acknowledgments}
We would like to thank Prof. P. K. Mohanty for useful discussions. DD's research is partially supported by the grant SR/S2/JCB-24/2006 by the Department of Science and Technology, India.
\end{acknowledgments}

\end{document}